\documentclass[a4paper, superscriptaddress, amsfonts, amssymb, amsmath, reprint, showkeys, nofootinbib, twoside]{revtex4-1}
\usepackage[english]{babel}
\usepackage[utf8]{inputenc}
\usepackage{xcolor}
\usepackage{bm}
\usepackage{graphicx}
\usepackage{float}
\usepackage{amsmath}
\usepackage{amsfonts,amssymb,amsthm}
\usepackage[colorinlistoftodos, color=green!40, prependcaption]{todonotes}
\usepackage{amsthm}
\usepackage{mathtools}
\usepackage{physics}
\usepackage{xcolor}
\usepackage{graphicx}
\usepackage[left=23mm,right=13mm,top=35mm,columnsep=15pt]{geometry} 
\usepackage{adjustbox}
\usepackage{placeins}
\usepackage[T1]{fontenc}
\usepackage{lipsum}
\usepackage{csquotes}
\usepackage[pdftex, pdftitle={Article}, pdfauthor={Author}]{hyperref} 

\begin{document}
\title{Investigating High-Order Behaviors in Multivariate Cardiovascular Interactions via Nonlinear Prediction and Information-Theoretic Tools}

\author{Chiara Barà}
    \affiliation{Institute of Intelligent Industrial Technologies and Systems for Advanced Manufacturing, National Research Council, Milan 20133, Italy}
\author{Yuri Antonacci}
    \affiliation{Biosignals and Information Theory Laboratory, Department of Engineering, University of Palermo, 90128 Palermo, Italy}
\author{Laura Sparacino}
    \affiliation{Biosignals and Information Theory Laboratory, Department of Engineering, University of Palermo, 90128 Palermo, Italy}
\author{Helder Pinto}
    \affiliation{Departamento de Matemática, Faculdade de Ciências, Universidade do Porto, 4169-007 Porto, Portugal}   
\author{Michal Javorka}
    \affiliation{Department of Physiology, Comenius University in Bratislava, Jessenius Faculty of Medicine, 036 01 Martin, Slovakia}
\author{Sebastiano Stramaglia}
    \affiliation{Department of Physics, University of Bari Aldo Moro, 70126 Bari, and INFN, Sezione di Bari, Italy}
\author{Luca Faes}
    \email[Correspondence email address: ]{luca.faes@unipa.it}
    \affiliation{Biosignals and Information Theory Laboratory, Department of Engineering, University of Palermo, 90128 Palermo, Italy}

\begin{abstract}
\textit{Objective}: Assessing the synergistic high-order behaviors (HOBs) that emerge from underlying structural mechanisms is crucial to characterize complex systems. This work leverages the combined use of predictability and information-theoretic measures to detect and quantify HOBs in synthetic and physiological network systems.
\textit{Methods}: After providing formal definitions of mechanisms and behaviors in a complex system, measures of statistical synergy are defined as the whole-minus-sum (WMS) excess of mutual predictability ($\Delta_\textrm{MP}$) or mutual information ($\Delta_\textrm{MI}$) observed when considering the system as a whole rather than as a combination of its units. The two measures are computed using model-free methods based on nonlinear prediction and entropy estimation.
\textit{Results}: The application to simulated linear Gaussian systems and nonlinear deterministic and stochastic dynamic systems shows that $\Delta_\textrm{MP}$ tends to vanish for target variables influenced by additive effects of single independent source variables and is positive in the presence of group interactions between sources, while $\Delta_\textrm{MI}$ exhibits a higher propensity to display positive values. Then, the analysis of physiological variables shows significant values of $\Delta_\textrm{MI}$ when investigating the additive effect of systolic and diastolic arterial pressure on mean arterial pressure, and of both $\Delta_\textrm{MP}$ and $\Delta_\textrm{MI}$ when assessing  how diastolic pressure is modulated by pre-ejection period and left-ventricular ejection time.
\textit{Conclusion}: HOBs can be more clearly identified by information-theoretic WMS measures, while prediction WMS measures appear more sensitive to synergy arising from the governing rules of the system analyzed rather than from pure statistical dependencies.
\textit{Significance}: Quantifying HOBs through WMS measures sensitive to complex structural mechanisms can provide new biomarkers to assess physio-pathological alterations of cardiovascular networks.
\end{abstract}


\maketitle

\section{Introduction}

Physiological dynamics are the result of complex interactions arising between subparts of the human body at different scales, ranging from small cells to entire organ systems \cite{barabasi2013network}. In this scenario, networks have become a paradigmatic tool to describe the architecture of interconnected physiological systems, with representations spanning from the structural links within each physiological system to the functional activity and connectivity of different systems \cite{ivanov2021new}. The field of Network Physiology \cite{ivanov2021new} is commonly investigated exploiting tools developed in the broader field of Network Science \cite{boccaletti2006complex, barabasi2013network} to assess links between central and peripheral organ systems, e.g., for the analysis of brain-heart interactions \cite{candia2025measures}, or between parts of the same physiological system, e.g., to characterize cardiovascular interactions \cite{lehnertz2020human}. Nevertheless, it is nowadays widely acknowledged that a comprehensive understanding of real-world network systems necessitates a shift from simplistic representations that merely depict the nodes and the links between pairs of them, to more sophisticated descriptions incorporating interactions among more than two nodes. These interactions, commonly denoted as \textit{high-order interactions} (HOIs), have become the focus of several recent investigations within the field of Network Physiology \cite{faes2022new, mijatovic2024assessing, bara2025direct}.

Despite the longstanding recognition of group interactions within systems with a network architecture \cite{anderson1972more, atkin1974mathematical, berge1984hypergraphs}, systematic analyses of HOIs have only recently gained popularity in the study of complex network systems \cite{battiston2020networks}, thanks to two distinct lines of inquiry: the study of the structure of group interactions, with emphasis on the detection of the rules governing the causal evolution of the system, typically referred to as \textit{high-order mechanisms} (HOMs) \cite{malizia2024reconstructing}, and the statistical assessment of patterns of activity that can be explained by the system as a whole but not by any of its parts considered separately, denoted as \textit{high-order behaviors} (HOBs) \cite{rosas2019quantifying, faes2022new}. Crucially, mechanisms and behaviors provide complementary but closely related perspectives on the description of high-order phenomena. While it is established that HOMs and HOBs are interrelated, their interplay is far from trivial \cite{rosas2022disentangling, robiglio2025synergistic}: simple structures can generate complex HOBs, and even when complex structures are present and give rise to HOMs, they may produce different types of HOBs \cite{robiglio2025synergistic, marinazzo2025behaviors}. Consequently, identifying behavioral signatures can help constrain the set of possible underlying mechanisms, but it does not guarantee that these mechanisms are fully captured \cite{marinazzo2025behaviors}.

One reason behind the difficulty of assessing the interplay between the two forms of HOIs described above is the fact that, while HOMs can be clearly formulated through mathematical laws involving beyond-pairwise interactions, the definition of HOBs is less unequivocal. In general, HOBs are related to the concept of statistical synergy, according to which the system behavior can be better predicted from the joint consideration of all its units rather than using the units in isolation \cite{rosas2019quantifying, marinazzo2025behaviors, varley2025topology}. Several approaches exist to operationalize this concept based on nonlinear prediction \cite{porta2017quantifying} or information theory \cite{mcgill1954multivariate}. In this context, the present work formalizes two different approaches to quantify HOBs from network systems, and compares them with each other first in simulated systems with known underlying mechanisms to investigate the interplay between HOMs and HOBs, and then in a cardiovascular dataset to assess the presence of HOBs in simple physiological networks whose underlying interactions are established or well-acknowledged in the literature.

Specifically, modelling the behavior of a network system in terms of random variables, we quantify synergy as the difference in predictive power or in entropy, measured between a target variable and two source variables when the sources are taken jointly to describe the target, compared to when they are taken separately. The resulting ``whole-minus-sum'' (WMS) measures of HOBs are implemented in practice employing nearest-neighbor techniques for nonlinear prediction or entropy estimation. As regards the validation on the simulated settings, these measures are first compared in a theoretical analysis of Gaussian systems characterized by the absence of HOMs and then in empirical analyses of nonlinear deterministic and stochastic dynamic systems with and without HOMs. Finally, the method is applied to cardiovascular interactions to assess HOBs in the description of the variability of mean arterial pressure (MAP) from systolic (SAP) and diastolic arterial pressure (DAP), and of the variability of DAP from cardiac pre-ejection period (PEP) and left-ventricular ejection time (LVET), measured in a resting state and during a postural stress task evoking changes in the neuro-autonomic modulation of these variables.

\section{Methods}\label{Methods}

\subsection{Definitions of high-order mechanisms and behaviors} \label{sec:defHOMHOB}

Let us consider a network system $\mathcal{S}=\{\mathcal{Y},\mathcal{X}_1,\ldots,\mathcal{X}_N\}$ in which the activities of the single nodes are described by the target random variable $Y$ and the vector of the source variables $\boldsymbol{X}=[X_1 \cdots X_N]$.
HOMs are denoted by the presence of group interactions between the sources and the target that cannot be modeled simply using laws involving their pairwise relations \cite{rosas2022disentangling, robiglio2025synergistic}. Accordingly, in a system \textit{without} HOMs the target results from a noisy function that can be written as a sum of functions each depending on only one source, i.e., $Y=f(\boldsymbol{X})=f_1(X_1)+\cdots +f_N(X_N)+U$, where $f_i$ are general nonlinear functions and $U\perp \boldsymbol{X}$ is a variable capturing the portion of $Y$ unrelated to $X_i$, $i=1,\ldots,N$. When this generalized additive model cannot fully represent the relation between $Y$ and  $\boldsymbol{X}$, e.g., when cross-terms $X_i\cdot X_j$ appear in $f(\cdot)$, the system is said to possess HOMs w.r.t. the target $Y$.

Differently, a system is said to display HOBs when the statistical dependence of the target on the sources cannot be attributed to any source on its own, but emerges only from the way the sources interact together \cite{mcgill1954multivariate, rosas2022disentangling}. Accordingly, given a measure $C(\cdot;\cdot)$ quantifying the statistical coupling between pairs of (possibly vector) variables, the system is said to possess HOBs w.r.t. the target $Y$ when the following synergistic relation holds: $C(Y;\boldsymbol{X})>\sum_{i=1}^NC(Y;X_i)$. This definition implements the WMS concept whereby synergy arises if the system as a whole is more than the sum of its parts.

Without loss of generality, in the following we will considered only three-node systems ($N=2$); theoretical and computational issues arising for large systems are discussed in Sect. \ref{sec_disc}.

\subsection{Quantifying high-order behaviors through predictability and information-theoretic measures} \label{delta_measures}

In this section, the concept of synergy is operationalized exploiting both nonlinear prediction and information theory, defining two distinct WMS measures which quantify how much the activity at a target node in a network system is explained better considering the combined influences of all other nodes than examining their effect in isolation.

\subsubsection{Predictability measure of HOBs} \label{HOPred}
Within the framework of predictability \cite{faes2016predictability}, the synergistic interaction of the sources in influencing the target can be investigated by evaluating how the prediction of the target variable is improved when considering the source variables together rather than separately. In the two-sources case, this is achieved by decomposing the variance of the target into terms related to the residual uncertainty that remains when one or both sources are known.
This concept is quantified by the measure of \textit{mutual predictability} (MP), which is defined, when prediction is performed using only one source variable $X_i$ ($i=1,2$), as the difference between the variance of the target, $\sigma_Y^2$, and its residual variance, $\sigma_{Y|X_i}^2$, i.e., $\lambda^2(Y;X_i) = \sigma_Y^2 - \sigma_{Y|X_i}^2$.
Considering the MP terms between the target and the sources taken both jointly, $\lambda^2(Y;\boldsymbol{X})$, and separately, $\lambda^2(Y;X_1)$ and $\lambda^2(Y;X_2)$, the WMS measure of synergy is defined as 
\begin{equation}\label{delta_MP}
\begin{aligned} 
    \Delta_\textrm{MP}(Y;\boldsymbol{X}) &= \lambda^2(Y;\boldsymbol{X}) - \lambda^2(Y;X_1) - \lambda^2(Y;X_2) \\
    &=\sigma_{Y|X_1}^2+\sigma_{Y|X_2}^2-\sigma_{Y}^2-\sigma_{Y|\boldsymbol{X}}^2.
\end{aligned}
\end{equation}
The measure (\ref{delta_MP}) is related to the interaction predictability index defined in \cite{porta2017quantifying}. It takes positive values when, thanks to the synergistic interplay between the two sources, the target is better predicted using both sources together rather than in isolation.

\subsubsection{Information measure of HOBs} \label{HOInfo}
An information-theoretic quantification of the synergistic interactions between two sources and a target variable can be obtained by evaluating the amount of information that the target shares with the sources considered jointly, beyond the information it shares with each source considered individually. This concept is implemented by combining terms representative of the \textit{mutual information} (MI) shared between the target and the sources. MI is defined, when considering only one source variable $X_i$, as the difference between the entropy of the target, $H(Y)$, and its entropy conditioned on the source, $H(Y|X_i)$, i.e., $I(Y;X_i) = H(Y) - H(Y|X_i)$. Considering the MI terms between the target variable and the sources together, $I(Y;\boldsymbol{X})$, and separately, $I(Y;X_1)$ and $I(Y;X_2)$, the WMS measure of synergy is defined as
\begin{equation} \label{delta_MI}
\begin{aligned}
    \Delta_\textrm{MI}(Y;\boldsymbol{X}) &= I(Y;\boldsymbol{X})-I(Y;X_1) - I(Y;X_2) \\
    &= H(Y|X_1) + H(Y|X_2) - H(Y) - H(Y|\boldsymbol{X}).
\end{aligned}
\end{equation}
This measure corresponds with the well-known interaction information \cite{mcgill1954multivariate} and assumes positive values when the information shared by the target and the joint sources is greater than the sum of the information shared by the target and the sources separately, thus indicating the presence of a synergistic interplay.

\subsection{Practical computation} \label{pract}
Considering a realization of length $N$ of the target variable $Y$, i.e., $y=\{y_1,\dots,y_N\}$, and of the source variables $X_1$ and $X_2$, i.e., $x_1=\{x_{1,1},\dots,x_{1,N}\}$ and $x_2=\{x_{2,1},\dots,x_{2,N}\}$, the estimation of predictability and information measures is achieved through the use of model-free approaches, which facilitate the characterization of nonlinearities possibly involved in the analyzed interactions. This constitutes a crucial aspect in our analysis, as the employment of linear approaches would be unable to identify nonlinear interactions that are present in systems characterized by HOMs and that can also be present in systems characterized by HOBs.
Specifically, MP terms are quantified through a nonlinear approach based on \textit{k}-nearest neighbor local linear approximation (Sect. \ref{MP_estimation}) \cite{farmer1987predicting}, while the MI terms are evaluated by using the model-free nearest neighbor estimation strategy (Sect. \ref{MI_estimation}) \cite{kraskov2004estimating}. 

\subsubsection{MP estimation} \label{MP_estimation}

The nonlinear approach used here to estimate predictability measures is based on the idea that the interdependencies among target and sources can be locally approximated as linear even if their global relation is assumed to be nonlinear \cite{farmer1987predicting}. 
The approach is used to compute the three MP terms appearing in (\ref{delta_MP}) to quantify $\Delta_\textrm{MP}$, i.e., $\lambda^2(Y;\boldsymbol{X})$, $\lambda^2(Y;X_1)$, and $\lambda^2(Y;X_2)$. Therefore, this is presented in the following for a generic source variable $X \in \{X_1, X_2, \mathbf{X}\}$.

For each sample of the target variable $y_n$, a neighbor search is performed in the source space $\{X\}$ by identifying the \textit{k} nearest neighbors with lowest Chebyshev distance from the source sample $x_n$, i.e., $z_n = [z_{n_1},\dots,z_{n_k}]$ being $z_{n_i}$, $i=1,\dots,k$, the $i$-th neighbor of $x_n$. These neighbor samples are used to write a system of $k$ equations which linearly approximate the dependence of the target variable on the source, i.e., $y_{n_i} = a_i z_{n_i} + u_{n_i}$, where $u_{n_i}$ is the prediction error. The least squares approach is used to solve the linear system and the estimated vector coefficient $\hat{A}=[\hat{a}_1,\dots,\hat{a}_k]$ is then used to predict the target sample, i.e., $\hat{y}_n = \hat{A}z_n$. Moreover, the current sample of the prediction error is obtained as $\hat{u}_n=y_n-\hat{y}_n$ \cite{farmer1987predicting}. In this way, it is possible to obtain a value of the prediction error for each sample of $Y$ and $X$ and thus to compute its variance. This latter corresponds to the partial variance of $Y$ that cannot be explained by $X$, i.e., $\sigma^2_U=\sigma^2_{Y|X}$, and is subtracted from the variance of the target variable to compute the MP term $\lambda^2(Y;X) = \sigma_Y^2-\sigma^2_U$.

In this work, the optimization procedure introduced in \cite{erla2011k} was used to set the number of neighbors \textit{k} to be used for each of the samples and based on the maximization of the predictability of the target variable given the source. Specifically, the variance of the prediction error $\sigma^2_U$ was evaluated for values of $k$ ranging from very local prediction ($k = 2$) to global prediction ($k = N$) and then the value of $k$ corresponding to the lowest $\sigma^2_U$ is considered optimal and used to find the MP term.

\subsubsection{MI estimation} \label{MI_estimation}

The model-free nearest-neighbor estimation approach was adopted for the computation of the MI terms needed to obtain $\Delta_\textrm{MI}$ (see Eq. \ref{delta_MI}) \cite{kozachenko1987sample}. Specifically, accounting for the estimation bias deriving from the combination of information terms computed on variables of different dimensions, the formulation introduced in \cite{kraskov2004estimating} was employed .

Being the joint space of the target variable and both sources, i.e., $[Y,\boldsymbol{X}]$, the highest-dimensional observed space, $k$ neighbors are searched for each sample of the joint variable to compute $I(Y;\boldsymbol{X})$ and the projected range search used in the lower dimensional spaces to compute $I(Y;X_1)$ and $I(Y;X_2)$. Specifically, the formulas $I(Y;\boldsymbol{X}) = \psi(k)+\psi(N)-\langle \psi(N_Y+1) + \psi(N_{\boldsymbol{X}}+1) \rangle$ and $I(Y;X_i) = \psi(N)+\langle \psi(N_{YX_i}+1) -\psi(N_Y+1) - \psi(N_{X_i}+1)\rangle$, for $i=\{1,2\}$, are employed, where $\langle \cdot \rangle$ represents the average across the $N$ realizations of the variables and, being $\epsilon_{n,k}$ the maximum distance from the generic observation of $[Y,\boldsymbol{X}]$ to its $k$-th neighbor, with $N_Y$, $N_{\boldsymbol{X}}$, $N_{YX_i}$ and $N_{X_i}$ representing the number of samples in the spaces $\{Y\}$, $\{\boldsymbol{X}\}$, $\{Y \: X_i\}$, and $\{X_i\}$ with distance lower than $\epsilon_{n,k}$ from their generic observation, respectively. 
The number of neighbors $k$ was fixed to 5 according to common choice in literature, which enables a trade-off between the bias and variance of the estimated quantities \cite{bara2025partial}.

\subsection{Statistical significance} \label{surr}

A surrogate data analysis approach was used to assess the significance of the HOI measures evaluated through the predictability and information approaches, i.e., $\Delta_\textrm{MP}$ and $\Delta_\textrm{MI}$. To preserve the statistical structure of each individual variable while disrupting the correlations between the sources and the target -- both individually and jointly -- we applied a circular-shift surrogate strategy. Specifically, each source was shifted by a different number of samples, while the target remained unchanged \cite{porta2017quantifying, andrzejak2003bivariate}. The shift magnitude was randomly selected, excluding very small values that would be insufficient to break the correlation between the target and the sources. The samples shifted beyond the end of the original source data were wrapped around to the beginning of the surrogate series. The procedure was repeated $N_s$ times, and the considered measure was computed at each iteration, resulting in a distribution. To test for the presence of synergistic interactions (i.e., $\Delta > 0$), the $100(1-\alpha)$-th percentile of this distribution was extracted and compared with the original value. Statistical significance was established when the measure exceeded the extracted threshold. For the computation of the $\Delta_\textrm{MP}$ terms on surrogate data, the optimal values of $k$ found on the original ones for the computation of $\lambda^2(Y;\boldsymbol{X})$, $\lambda^2(Y;X_1)$, and $\lambda^2(Y;X_2)$, were employed. 

\section{Validation on simulations} \label{sect:simulations}

This section investigates the behavior of the WMS measures introduced in Sect. \ref{delta_measures} in investigating HOBs in five different simulated scenarios: i) a linear Gaussian system; ii) coupled logistic maps; iii) coupled Hénon maps; iv) nonlinear autoregressive process; and v) nonlinear mix of linear autoregressive processes. These deterministic and stochastic processes are mathematically well-defined, so their characterization in terms of HOMs can be established according to the definitions given in Sect. \ref{sec:defHOMHOB}. Specifically, since in linear Gaussian, coupled logistic and nonlinear stochastic processes the target variable depends on each source separately via additive functions, these systems do not present HOMs. The analyzed coupled Hénon maps and nonlinearly mixed stochastic processes, on the other hand, display a target variable related to the source variables through cross-terms, and thus present HOMs.

For the linear Gaussian system, a theoretical analysis was performed providing the exact mathematical derivation of the WMS measures based on MP and MI. In the remaining simulations, an empirical analysis was conducted by generating fifty realizations of length $N = 1000$ for each system and computing the WMS measures  for all the realizations; the statistical significance of $\Delta_\textrm{MP}$ and $\Delta_\textrm{MI}$ was assessed for each realization by using one-hundred surrogate data, generated by randomly shifting the source variables of at least 50 samples, and fixing a significance threshold of $\alpha=0.05$.

Note that, being all simulated systems except the linear Gaussian dynamic, the target and source variables needs to be chosen sampling the dynamic processes at a given time step; here, analyses were performed taking $Y_n$ as the target and $X_{1,n-1}$, $X_{2,n-1}$ as the sources (see Eqs. (\ref{Y_coupl_log}, \ref{Y_coup_henon}, \ref{Y_mixed}, \ref{Y_NAR})).

\subsection{Linear Gaussian system} \label{linear_Gaussian}

The first analyzed system is composed of three variables, $Y$, $X_1$ and $X_2$ with joint Gaussian distribution. Without loss of generality, we set zero mean and unit variance for all the variables. Since Gaussian systems are fully described by linear regression models \cite{barrett2010multivariate}, the MP and MI terms needed to compute the WMS measures (\ref{delta_MP}) and (\ref{delta_MI}) can be derived from the models relating the target variable to each individual source, $Y = aX_1+U_1$, $Y = bX_2+U_2$, and to both sources, $Y = \bar{a}X_1+\bar{b}X_2+U$, where $U_1$, $U_2$ and $U$ are the Gaussian residual variables unrelated to the predictors.
From these models, it is possible to show that the variance of the residuals can be written, in this case with $\sigma^2_Y=\sigma^2_{X_1}=\sigma^2_{X_2}=1$, as a function of the elements of the covariance matrix of the joint vector variable $[Y X_1 X_2]$ \cite{barrett2015exploration}:
\begin{subequations} \label{sigmaGauss}
    \begin{alignat}{2}
        \sigma^2_{U_i} &=\sigma^2_{Y|X_i}= 1-r^2_i, i=1,2\\
        \sigma^2_{U} &=\sigma^2_{Y|\boldsymbol{X}}= \frac{1-(r_1^2+r_2^2+r_{12}^2)+2r_1r_2r_{12}}{1-r^2_{12}},
    \end{alignat}
\end{subequations}
where $r_1=\mathbb{E}[YX_1]$, $r_2=\mathbb{E}[YX_2]$ and $r_{12}=\mathbb{E}[X_1X_2]$ are the pairwise correlations between the variables.
Then, the WMS predictability measure is obtained substituting (\ref{sigmaGauss}) in (\ref{delta_MP}):
\begin{equation}\label{delta_MP_Gauss}
    \Delta_\textrm{MP}(Y;\boldsymbol{X})=\frac{(r_1^2+r_2^2)r^2_{12}-2r_1r_2r_{12}}{1-r^2_{12}}.
\end{equation}
The WMS information measure is obtained considering that the MI between target and source variables become, for jointly Gaussian variables, $I(Y;X_i)=\frac{1}{2}\log\frac{\sigma^2_Y}{\sigma^2_{U_i}}$ and $I(Y;\boldsymbol{X})=\frac{1}{2}\log\frac{\sigma^2_Y}{\sigma^2_{U}}$, and substituting these MI terms in (\ref{delta_MI}):
\begin{equation}\label{delta_MP_Gauss}
    \Delta_\textrm{MI}(Y;\boldsymbol{X})=\frac{1}{2}\log \frac{(1-r_1^2)(1-r_2^2)(1-r^2_{12})}{1-(r_1^2+r_2^2+r_{12}^2)+2r_1r_2r_{12}}.
\end{equation}

\begin{figure}
    \centering
    \includegraphics[width=.49\textwidth]{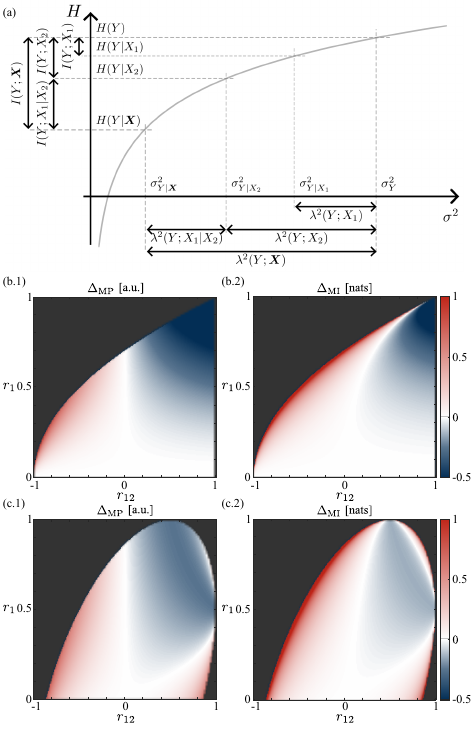}
    \caption{Detection of HOBs in linear Gaussian systems. (a) Graphical representation of the derivation for Gaussian variables (target $Y$, sources $X_1$, $X_2$) of the predictability and information-theoretic WMS measures defined in Eqs. (\ref{delta_MP}) and (\ref{delta_MI}); note that, if $X_1 \perp X_2$, $\lambda^2_{Y;X_1|X_2}=\lambda^2_{Y;X_1}$ and thus $\Delta_{\textrm{MP}}(Y;\boldsymbol{X})=0$, but $I(Y;X_1|X_2)>I(Y;X_1)$ and thus $\Delta_{\textrm{MI}}(Y;\boldsymbol{X})>0$. (b,c) Color-coded values of $\Delta_{\textrm{MP}}(Y;\boldsymbol{X})$ and $\Delta_{\textrm{MI}}(Y;\boldsymbol{X})$ at varying the strength of the link between the sources ($r_{12}$) and between the target and each source, either balanced ($r_1=r_2$, (b)) or unbalanced ($r_1$, with fixed $r_2=0.5$, (c)).}
    \label{fig:demonstration}
\end{figure}

The above reported analytical formulations allow us to determine the exact values of the WMS measures of synergy based on predictability and entropy as a function of the correlations between pairs of Gaussian variables composing the observed triplet.
A first important observation is that when $r_{12}=0$, we have $\Delta_\textrm{MP}=0$ and $\Delta_\textrm{MI}>0$, indicating that if the sources $X_1$ and $X_2$ are uncorrelated the WMS entropy measure detects synergistic HOBs, while the WMS predictability measure does not. This result is a consequence of the fact that for independent predictors the predictive power is additive and thus $\Delta_\textrm{MP}=0$ by Eq. (\ref{delta_MP}); on the other hand, as depicted in Fig. \ref{fig:demonstration}(a), the concave property of the logarithm makes $\Delta_\textrm{MI}>0$.

To investigate how HOBs emerge from the correlation structure of the system, we performed a systematic analysis of the measures $\Delta_\textrm{MP}$ and $\Delta_\textrm{MI}$ at varying the pairwise links between variables. Fig. \ref{fig:demonstration}(b) shows that, in the presence of equal-strength links between the target and each source ($r_1=r_2$), only the entropy-based measure can take positive values when the sources are positively correlated ($r_{12}>0$) while both measures can be positive when the sources are anti-correlated ($r_{12}<0$). When the target-source correlations are unbalanced (Fig. \ref{fig:demonstration}(c)), both measures can be positive for anti-correlated sources but also for positively correlated sources if the target tends to be uncorrelated with one of them; the entropy-based measure displays positive values for a wider range of parameter combinations (Fig. \ref{fig:demonstration}(c)).

Overall, this simulation evidences that HOBs can emerge even for simple linear systems where the generating equations exclude the presence of HOMs, and that the entropy-based WMS measure shows a higher propensity to display HOBs.

\subsection{Nonlinear deterministic system without HOMs} \label{coupled_logistic_maps}

In the first empirical simulation (Sim 1), two logistic maps in the chaotic regime are coupled linearly to generate a discrete-time deterministic nonlinear system defined by the equations \cite{kantz2003nonlinear, sugihara2012detecting}:
\begin{subequations}
    \begin{alignat}{3}
        X_{1,n} &= 3.9 X_{1,n-1}(1-X_{1,n-1}),\\
        X_{2,n} &= 3.7 X_{2,n-1}(1-X_{2,n-1}),\\
        Y_{n} &= c_1 X_{1,n-1}+c_2 X_{2,n-1} + 0.1 Y_{n-1}, \label{Y_coupl_log}
    \end{alignat}
\end{subequations}
where $c_1=0.9$ and $c_2=0.7$ are the parameters controlling the linear influence of the processes $X_1$ and $X_2$ on $Y$, respectively. Initial random conditions were selected from uniform distribution in the range $[0 \:\: 1]$ and a transient period of 500 samples discarded for each realization. 

As shown in Fig. \ref{fig_sim}, this setting results in high and always significant $\Delta_\textrm{MI}$ values and in values of $\Delta_\textrm{MP}$ very close to zero (about half of the realizations detect significant synergy). These results reproduce empirically the results of the theoretical simulation in Sect. \ref{sec:defHOMHOB} showing that, in this nonlinear system with independent non-Gaussian sources and in the absence of HOMs, HOBs are clearly detected by entropy-based WMS measure but not by the prediction-based one.

\subsection{Nonlinear deterministic system with HOMs} \label{coupled_henon}
In the second empirical simulation (Sim 2), we study a nonlinear deterministic system composed by three coupled Hénon maps, defined by the equations \cite{kugiumtzis2013direct, faes2015estimating}:
\begin{subequations}
    \begin{alignat}{2}
        X_{1,n} &= 1.4 - X^2_{1,n-1} + 0.3 X_{1,n-2}, \\
        X_{2,n} &= 1.4 - X^2_{2,n-1} + 0.3 X_{2,n-2}, \\
        Y_{n} &= 1.4 - 0.25(X_{1,n-1}+X_{2,n-1})^2+0.3Y_{n-1}. \label{Y_coup_henon}
    \end{alignat}
\end{subequations}
Again, for each realization, initial random conditions were fixed by considering uniformly distributed values in the range $[0 \:\: 1]$ and the initial transient period of 500 samples discarded. 

Fig. \ref{fig_sim} shows how both $\Delta_\textrm{MP}$ and $\Delta_\textrm{MI}$ take high positive values which are statistically significant in all realizations of the simulation. These results indicate that, in this nonlinear system with independent non-Gaussian sources and in the presence of HOMs, both the prediction-based and the entropy-based WMS measures clearly detect the presence of HOBs.

\subsection{Nonlinear stochastic system without HOMs} \label{linear_stochastic} 
The third simulation (Sim 3) realizes the nonlinear first-order vector autoregressive process defined by the stochastic equations \cite{gourevitch2006linear, kugiumtzis2013direct}:
\begin{subequations}
    \begin{alignat}{3}
        X_{1,n} &= 3.4 X_{1,n-1}(1-X^2_{1,n-1})e^{-X^2_{1,n-1}}+0.1e_{1,n},\\
        X_{2,n} &= 3.4 X_{2,n-1}(1-X^2_{2,n-1})e^{-X^2_{2,n-1}}\\
            &+0.5X_{1,n-1}X_{2,n-1}+0.1e_{2,t}, \nonumber\\
        Y_{n} &= 3.4 Y_{n-1}(1-Y^2_{n-1})e^{-Y^2_{n-1}}+c_1X^2_{1,n-1} \label{Y_mixed}\\ 
            &+c_2 X_{2,n-1}+0.1e_{3,t}, \nonumber 
    \end{alignat}
\end{subequations}
where the coefficients $c_1=0.5$ and $c_2=0.3$ set the strength of the additive effect of the sources $X_1$ and $X_2$ on the target $Y$, respectively, and $e_i$, $i=1,\dots,3$ are uncorrelated white Gaussian noises with zero mean and unit variance. Here, realizations of the simulations were obtained drawing random samples for the noise inputs. 

The results depicted in Fig. \ref{fig_sim} highlight a similar situation than that described in Sect. \ref{coupled_logistic_maps}, with high and statistically significant values of $\Delta_\textrm{MI}$ and low and barely significant values of $\Delta_\textrm{MP}$. The larger dispersion of $\Delta_\textrm{MP}$ might be due to the correlation between the sources which could possibly induce a detection of HOBs also with the prediction-based method.

\subsection{Nonlinear stochastic system with HOMs} \label{NAR}
The last simulation (Sim 4) considers a system with two autoregressive processes $X_1$ and $X_2$ that  interact nonlinearly to determine a third process $Y$, according to \cite{kilian2017structural, terasvirta2010modelling}:
\begin{subequations}
    \begin{alignat}{3}
        X_{1,n} &= 0.8 X_{1,n-1}+0.2e_{1,n},\\
        X_{2,n} &= 0.6 X_{2,n-1}+0.2e_{2,n},\\
        Y_{n} &= 0.5 Y_{n-1}+c(X_{1,n-1}X_{2,n-1})+0.2e_{3,t}, \label{Y_NAR}
    \end{alignat}
\end{subequations}
being $c=2$ the coupling parameter, and $e_i$, $i=1,\dots,3$, uncorrelated Gaussian noises with zero mean and unit variance. Again, different realizations of the processes where obtained drawing random samples for the inputs and iterating the system equations. 

\begin{figure}
    \centering
    \includegraphics[width=.49\textwidth]{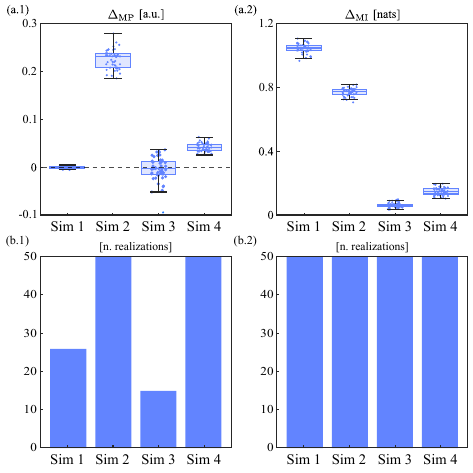}
    \caption{Detection of HOBs in nonlinear deterministic and stochastic systems with and without HOMs. Panels depict boxplots with individual values of the predictability (a.1) and information-theoretic (a.2) WMS measures, and number of realizations for which the measures were detected as statistically significant (b.1, b.2), for the simulations of coupled logistic maps (Sim 1), coupled Hénon maps (Sim 2), nonlinear autoregressive processes (Sim 3), and nonlinearly coupled linear autoregressive processes (Sim 4).}
    \label{fig_sim}
\end{figure}

Fig. \ref{fig_sim} shows that, similarly to the system described in Sect. \ref{coupled_henon}, both $\Delta_\textrm{MP}$ and $\Delta_\textrm{MI}$ measures are positive and statistically significant in all realizations, documenting the existence of HOBs according to both prediction and entropy approaches.

\section{Application to physiological data} \label{appl}

The proposed frameworks of analysis were applied to investigate the presence of synergistic HOBs in two representative small-size networks underlying the short-term regulation of cardiovascular system. Specifically, the two proposed WMS measures were computed to investigate arterial pressure interactions, exploring the well-known linear relationship explaining how MAP is determined by SAP and DAP \cite{papaioannou2016mean}, as well as cardiovascular interactions, exploring the less studied nonlinear dependence between mechanical heart timing parameters of PEP and LVET in influencing DAP values \cite{cinelli1986relationship}. 

\subsection{Experimental protocol and data acquisition}
Part of a big dataset comprising cardiovascular and cardiorespiratory signals for investigating physiological response to stressors was employed \cite{javorka2018towards, svec2021short, iovino2024comparison}. Specifically, data acquired from a cohort of 84 young healthy subjects (50 females, age range 18.65$\pm$3.20 years) performing the head-up tilt maneuver were analyzed. 

Subjects underwent a two phases protocol consisting of a resting state period of 15 minutes (REST), during which they lied in a supine position, and of an orthostatic stress condition of 8 minutes (HUT), during which they were passively tilted to 45 degrees by means of a motorized bed. All subjects were asked to spontaneously breathe during the execution of the protocol. No signs of presyncope were evidenced for the whole cohort during HUT.

The recorded signals were the electrocardiogram (ECG), finger photoplethysmographic arterial blood pressure (BP, volume-clamp method) and impedance cardiography (ICG) signal.
The latter signal evaluates changes in the trans-thoracic impedance $\Delta Z$ as proportional to volume and blood flow velocity variations in the aorta, as well as changes in blood volume in the transthoracic region over time as the first mathematical derivative of $\Delta Z$, i.e., $dZ/dt$. All signals were synchronously acquired with a sampling frequency of 1 kHz.

The experimental protocol was approved by the ethical committee of the Jessenius Faculty of Medicine, Comenius University, Slovakia. All subjects or their legal representatives, in participants under 18 years of age, signed a written informed consent before the examination. Detailed information on experimental protocol and signal acquisition is provided in \cite{javorka2018towards, svec2021short}.

\subsection{Time series extraction and pre-processing}

Starting from the acquired signals, the following physiological beat-to-beat time series representative of the variability of different cardiovascular parameters were extracted. 
The $n^{\mathrm{th}}$ systolic and diastolic arterial pressure, $S_n$ and $D_n$, were obtained as the first maximum of the BP waveform measured after the $n^{\mathrm{th}}$ heartbeat detected from the ECG, and as the first minimum measured after the $(n+1)^{\mathrm{th}}$ heartbeat, respectively. The mean arterial pressure, $M_n$, was calculated as the average of the  BP waveform considered between $D_{n-1}$ and $D_n$.
Moeover, starting from the $dZ/dt$ waveform, PEP and LVET time series were obtained as follows. After the identification of the points B (first notch after $n$-th ECG R peak) and X (first significant minimum after $n^{\mathrm{th}}$ ECG R peak) corresponding to the opening and closure of the aortic valve, respectively, $P_n$ is obtained as the time difference between the $n^{\mathrm{th}}$ B point in the ICG derivative and the onset of $n^{\mathrm{th}}$ Q wave in the ECG signal, and $L_n$ as the time difference between the $n^{\mathrm{th}}$ X and B points in $dZ/dt$. 

For each subject and condition, visually stationary 300-samples time series were extracted after the stabilization of the physiological parameters, i.e., around 8 minutes after the beginning the REST phase, and to avoid the transient period due to the tilting procedure, i.e., around 3 minutes after the beginning of the HUT phase. The considered time series were preprocessed through a zero-phase highpass autoregressive filter with a cutoff frequency of 0.0156 Hz to remove slow trends. Moreover, outliers detected as samples falling above the 75-th percentile or below the 25-th percentile of the distribution representing the whole series were removed and replaced via the cubic spline interpolation approach.

\subsection{Data and statistical analysis}
High-order interactions within the two systems composed of the variables $\{M, S, D\}$ and $\{D, P, L\}$ were evaluated by using the WMS measures $\Delta_\textrm{MP}$ and $\Delta_\textrm{MI}$ considering the variables $M$ and $D$ as target, respectively, and the other two variables as sources. MP and MI measures were computed as detailed in Sect. \ref{pract}, and the statistical significance of the two WMS measures evaluated for each subject, condition, and physiological system through the surrogate data analysis approach described in Sect. \ref{surr} generating one-hundred testing data by randomly shifting the two source variables of at least 20 samples. 

Statistical differences between the two acquisition phases were assessed for both $\Delta_\textrm{MP}$ and $\Delta_\textrm{MI}$ in the two physiological systems via the paired nonparametric Wilcoxon signed rank test. Moreover, the Fisher's test of proportion was applied to statistically evaluate the variation in the proportion of significant synergistic interactions occurred in the two conditions for both measures and settings, as well as in the two setting for both measures during the same condition. A significant threshold of $\alpha=0.05$ was employed for all statistical tests.

\subsection{Results}
\begin{figure}
    \centering
    \includegraphics[width=.49\textwidth]{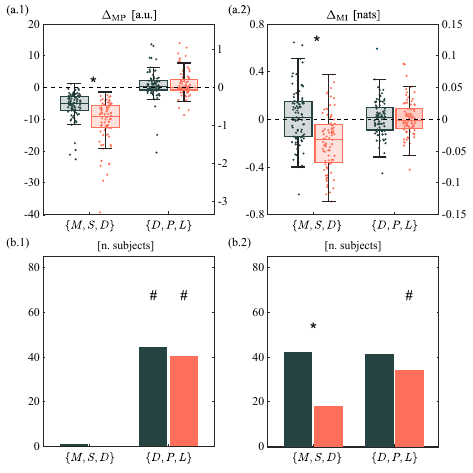}
    \caption{Detection of HOBs in cardiovascular interactions. Panels depict boxplots with individual values of the predictability (a.1) and information-theoretic (a.1) WMS measures, and number of realizations for which the measures were detected as statistically significant (b.1, b.2), for the two investigated physiological networks, i.e., $\{M, S, D\}$ obtained considering MAP (M) as target and SAP (S) and DAP as sources, and $\{D, P, L\}$ obtained considering DAP (D) as target and PEP (P) and LVET (L) as sources, at REST (bluish) and during HUT (pinkish). Statistical test in (a)): Wilcoxon signed rank test, * $p<0.05$, REST vs. HUT. Statistical tests in (b): Fisher's test, * $p<0.05$, REST vs. HUT; $\#$ $p<0.05$, $\{M, S, D\}$ vs. $\{D, P, L\}$.}
    \label{fig_appl}
\end{figure}

Figure \ref{fig_appl} shows the distributions across subjects of $\Delta_\textrm{MP}$ (panel \textit{a.1}) and $\Delta_\textrm{MI}$ (panel \textit{a.2}) measures considering the contribution of SAP and DAP to MAP (left) and of PEP and LVET to DAP (right), as well as the number of subjects for which the two WMS measures are significant (panels \textit{b}).

Overall, $\Delta_\textrm{MP}$ takes negative values when considering the $\{M, S, D\}$ system, with a significant decrease from REST to HUT and almost zero significance in both conditions. On the contrary, $\Delta_\textrm{MP}$ is mostly positive for the $\{D, P, L\}$ system, with values that are low but statistically significant for approximately half of the subjects during both REST and HUT. The Fisher test confirms the significance of this modulation across the two investigated systems. 

As regards  $\Delta_\textrm{MI}$, it reveals a slightly positive average value at REST in the system $\{M, S, D\}$ which significantly decreases during HUT leading to a prevalence of negative values; this decrease is  confirmed statistically by comparing the number of subjects for which the WMS measure is assessed as significant. On the other hand, the values of $\Delta_\textrm{MI}$ obtained for the system $\{D, P, L\}$ are mainly unvaried between conditions, with average close to zero and a number of statistically significant positive values close to half of the subjects in both conditions.
The Fisher test reveals a significantly higher number of subjects demonstrating synergistic interactions in the system $\{D, P, L\}$ compared to the system $\{M, S, D\}$ during HUT.

\section{Discussion} \label{sec_disc}

This study presents a thorough analysis of HOBs in complex network systems using nonlinear prediction and information-theoretic frameworks, aimed at investigating how the detection of functional HOIs depends on the  underlying structural HOMs. In particular, we examine how WMS measures of statistical high-order dependencies based on predictability or entropy functionals behave similarly or differently in the presence or absence of structural high-order effects in the generative equations.
First, this analysis was carried out on the benchmark scenario of linear Gaussian systems, which allows for the investigation of HOBs in the absence of HOMs. Then, it was extended on both nonlinear deterministic and stochastic systems displaying either additive source–target relationships or groupwise interactions, in accordance with the definition of HOMs as mechanisms arising from structural dependencies that cannot be decomposed into separate effects of individual sources. Lastly, the analysis was conducted on physiological variables descriptive of cardiovascular control systems, whose underlying relations could be hypothesized as they have been previously investigated in the literature.

Our analyses conducted in various simulation scenarios suggest that synergistic behavioral effects are quite commonly detected across different types of network systems, ranging from linear to nonlinear and from static to dynamic systems, as well as from deterministic to stochastic systems.
Specifically, we observed how HOBs emerge in systems characterized by groupwise interactions, thus defined by HOMs, as in the case of nonlinearly coupled Hènon maps and autoregressive processes, but also in systems determined by a simple additive effect of their parts, thus in systems without HOMs as in the case of linearly mixed logistic maps and nonlinear stochastic processes, and even in simple linear Gaussian systems.
Moreover, the comparison of the results achieved by using the WMS concept operationalized through the MP and MI measures provides important insights about the detection of HOBs, both in general and in relation to the presence of HOMs. First, we found that the measure $\Delta_\textrm{MI}$ is more effective than the measure $\Delta_\textrm{MP}$ in displaying synergistic HOBs, as it took positive values for more parameter combinations in the theoretical analysis of Gaussian systems (Fig. \ref{fig:demonstration}), and assumed statistically significant values for all the settings simulated in the empirical analysis of nonlinear systems (Fig. \ref{fig_sim}).
Furthermore, the full statistical significance of both $\Delta_\textrm{MP}$ and $\Delta_\textrm{MI}$ in the presence of HOMs (see Sim 2 and Sim 4 in Fig. \ref{fig_sim}) suggests the ability of the both the proposed WMS measures to detect HOBs evoked by HOMs. 
Nevertheless, we observed that the measure $\Delta_\textrm{MP}$ vanishes in the absence of HOMs when the sources are independent (e.g., for the linear Gaussian system when $r_{12}=0$, as well as in Sim 1 and Sim 3); this results confirms recent theoretical works \cite{barrett2015exploration, konig2024disentangling}.
However, $\Delta_\textrm{MP}$ turned out to be significantly positive even without HOMs when the sources are correlated (see Fig. \ref{fig:demonstration}(b.1,c.1)).
Therefore we conclude that, compared to $\Delta_\textrm{MI}$, $\Delta_\textrm{MP}$ is less sensitive to HOBs but more specific to HOMs, although its specificity is reduced by the correlation among the sources.
Overall, in line with the many studies employing information-theoretic approaches to investigate functional HOIs \cite{rosas2019quantifying, faes2022new, mijatovic2024assessing}, our observations suggest that $\Delta_\textrm{MI}$ should be preferred to assess HOBs, and that nonlinear prediction measures should be further investigated as a promising tool to dissect HOMs from empirical data \cite{konig2024disentangling, ontivero2025assessing}.

Along the same line of reasoning, the application of the two WMS measures on the two considered cardiovascular systems revealed how HOIs investigated in terms of behaviors can be related to mechanisms. Importantly, the physiological behaviors investigated in the two analyzed cardiovascular systems are clinically relevant: deeper knowledge about the relations between arterial pressure components, as well as with heart timing parameters reflecting the mechanical functioning of heart, could be of interest in the treatment of patients with stroke, head injury and hypertension \cite{papaioannou2016mean, cinelli1986relationship} and the evaluation other hemodynamic parameters \cite{svec2021short, cinelli1986relationship}.
When considering the system $\{D,P,L\}$ investigating the effects of PEP, reflecting the crucial time intervals between electrical depolarization and ejection of blood from the heart, and of LVET, reflecting the heart's phase of pumping blood in the aorta, on the diastolic arterial pressure, positive and quite significant values of both $\Delta_\textrm{MP}$ and $\Delta_\textrm{MI}$ were observed; this result, suggesting the possible presence of HOBs arising from HOMs, is supported by the nonlinear and non-additive underlying structural relation, and by the known correlation existing between $D_n$ and the ratio of $P_n$ and $L_n$\cite{cinelli1986relationship}. 
On the other hand, the negative $\Delta_\textrm{MP}$ values, together with the positive and more significant values for $\Delta_\textrm{MI}$ observed especially at rest for the system $\{M,S,D\}$, agree with the known presence of a linear and additive relation between SAP and DAP in determining MAP \cite{papaioannou2016mean}. In fact, when non-continuous blood pressure measurements are available, MAP values are usually estimated from SAP and DAP by through the relation $M_n = \frac{2}{3}S_n+\frac{1}{3}D_n$, which -- in accordance of our definition of HOMs -- suggests that the high-order characterization of the relation between arterial pressure components does not depend on a joint influence of SAP and DAP, but just from their additive effect. Nevertheless, it is interesting to note how the modulation between the two experimental
phases is coherent with the two WMS measures, revealing a decrease during HUT for the system $\{M,S,D\}$. This shift towards less synergistic interactions can be attributed to the impact of the mechanisms that act in response to orthostatic stress on arterial pressure dynamics – such as baroreflex mediated sympathetic activation – which in turn lead to alterations in arterial stiffness and vasomotion probably weakening the relation between the three arterial pressure components \cite{david2021short}. Moreover, in line with our observations in Sect. \ref{linear_Gaussian}, this decrease could be also be ascribed to the similar influence that such regulatory mechanisms have on SAP and DAP values strengthening their positive correlation. 

Our observations on both simulated and physiological data converge in identifying the predictability measure $\Delta_\textrm{MP}$ as a less sensitive quantity for detecting HOBs in absence of HOMs and, conversely, the information measure $\Delta_\textrm{MI}$ as a measures able to highlight the presence of HOBs, regardless of whether these are generated by hidden structural laws implying groupwise interactions or simple additive effects generating emergent statistical dependencies. 
Accordingly, using the WMS measures based on nonlinear predictability and information-theoretic frameworks in parallel presents a well-founded engineering approach for identifying the presence of HOBs and providing possible insights into the statistical or structural nature of the resulting high-order effects. The robustness of the achieved results, as well as the straightforward computation of $\Delta_\textrm{MP}$ and $\Delta_\textrm{MI}$ measures based on simple MP and MI terms, suggest the efficacy of our approach in the broad field of HOIs. In this regard, it is noteworthy to highlight that the methodologies currently used to explore these interactions are often based on a-priori assumptions about the system, computationally demanding approaches, or measures that are difficult to extend to high-dimensional systems \cite{williams2010nonnegative, torres2020simplicial, battiston2020networks}.
Nevertheless, it is also important to acknowledge the potential limitations of the proposed approach and the required improvements. First, estimation bias and variance must be considered, as these factors can obscure the results, particularly in the context of short-length data. Although not presented in the results, an analysis of the empirical simulations was performed using data lengths comparable to those of physiological data; despite similar trends in the WMS measures, higher estimation variance was observed, which strongly influenced the statistical significance of the measures and weakened the expected results. Second, although the concept of WMS is easy to grasp and implement, it conflates synergy and redundancy into a single metric, leading to an underestimation of the true high-order synergistic interactions. With this regard, future research should investigate approaches able to differentiate  synergistic and redundant components of interactions, e.g., the framework of Partial Information Decomposition \cite{williams2010nonnegative}, in characterizing HOBs arising from both structural or statistical dependencies. Finally, the feasibility of the proposed metrics should be deeply investigated in higher-dimensional systems due to problems related to bias compensation and the well-known problem of double counting redundancy affecting the WMS measures when the number of sources increases \cite{mediano2025toward}.

\section{Conclusion}

In conclusion, we propose using measures that implement the concept of WMS in analytical frameworks of nonlinear prediction and information theory to investigate HOBs in complex network systems. Our findings suggest that the information measure $\Delta_\textrm{MI}$ has a greater ability to observe higher-order effects when considering beyond-pairwise interactions and that the predictability measure $\Delta_\textrm{MP}$ is more sensitive in identifying HOBs arising from the presence of HOMs. Accordingly, the integrated perspective resulting from the proposed parallel approach rigorously exploits HOIs and possibly recognizes synergistic effects originating from system's underlying mechanisms or emergent behaviors.

Though preliminary, this work could enhance the ability to disambiguate the nature of the intricate interactions governing the physiological networks underlying the 
regulation of the homeostatic function of the human organism.
This can provide significant advancements in modelling and interpreting poorly understood physiological interactions, as well as in developing new biomarkers to assess physio-pathological alterations of cardiovascular networks.

\bibliography{biblio.bib}

\end{document}